\documentstyle[amssymb,prl,aps,epsfig]{revtex}
\newcommand{\bb}{\bibitem}

\begin{document}

\title{Stability and chaos around multipolar deformed bodies: A
 general relativistic approach}

 \author{Eduardo Gu\'eron\thanks{e-mail: gueron@ime.unicamp.br} 
 and
   Patricio S. Letelier\thanks{e-mail: letelier@ime.unicamp.br} }

\address{
Departamento de Matem\'atica Aplicada, Instituto de Matem\'atica,
Estat\'{\i}stica e Computa\c{c}\~ao Cient\'{\i}fica, Universidade 
Estadual de
Campinas, 13083-970, Campinas, SP, Brazil
}
 \maketitle
 
\begin{abstract}
The exact 
solution to the Einstein equations that represents  a static axially symmetric
   source deformed by an internal quadrupole is considered. By using 
 the Poincar\'e section method  we numerically 
study the  geodesic motion of test particles.
For the prolate quadrupolar deformations we  found chaotic motions contrary to the
oblate case where only regular motion is found. 
  We also  study 
 the metric that represents a rotating black hole deformed by a quadrupolar
term. This metric is obtained   as a two soliton solution
in the context of Belinsky--Zakharov inverse scattering method.
 The stability of geodesics
 depends strongly on the relative direction of the spin of the 
center of attraction and the test particle  angular momentum.
The rotation does not alter the regularity of geodesic
motions in the oblate case, i.e., the orbits in this  case
 remain regular.  We also employ the 
method of Lyapounov characteristic numbers
to examine the  stability of orbits evolving around  deformed 
 nonrotating centers of attraction. The typical time  to observe 
instability of orbits is analyzed.

\end{abstract}
\section{Introduction}

After the  visionary  work of Poincar\'e \cite{poincare} and 
 the KAM   (Kolmogorov,
 Arnol'd,  and Moser \cite{kam}) theory 
 it became well established that non-integrability and hence 
chaos is a general rather than exceptional manifestation in
 the context of dynamical systems, see for instance 
\cite{berry}. Given this ubiquitous fact,
 an important issue in astronomical   modeling is       the study in 
which extent in phase space chaoticity rises in models that
 are relevant to describe real systems and what are its 
consequences.

The adequate description of the gravitation field of an astrophysical
object has been an important subject in both relativistic and Newtonian 
gravity since their origin. The  particular case of the gravity associated 
to an axially symmetric body has played a central role in this discussion. 
Recently, Merrit \cite{merr} found,  from detailed  modeling of triaxial 
galaxies, that most of the galaxies must be nearly axisymmetric, either
 prolate or oblate. In Newtonian theory the gravitational potential of an 
axially symmetric body can be always represented by its usual expansion
 in terms of Legendre polynomials (zonal harmonics). 
The underlying theory in
 this case is 
the usual Newtonian Gravitation that for large masses and velocities 
is known to be less appropriate than  Einsteinian  General Relativity. 
In the late case the Newtonian potential is replaced by the 
spacetime metric and  Newton motion equations by geodesics. 
In General Relativity we have that the  solution of the vacuum
 Einstein equations associated
 to a static  axially symmetric body has a simple form with only
 two metric functions \cite{W} and one of them  admits an
 expansion in 
zonal harmonics. For  rotating axially symmetric bodies we have a metric
 with three functions and  two of them obey a sigma-model type of partial 
differential  equations for which there are known methods of
 solution \cite{cosg}.

The change of the particle motion equation and gravitational theory
can produce dramatic effects, for
 instance,  test particles moving in the presence of
 systems of masses that are integrable in Newtonian theory are chaotic
in General Relativity,  examples are: the fixed two body
  problem  \cite{cont,ssmae},
and particles moving in a monopolar 
center of attraction surrounded by a dipolar halo \cite{wldi}.
Also  gravitational waves, a non existing phenomenon in the
 Newtonian realm, can produce irregular motion of test particles
 orbiting around a static black hole \cite{bc,lwmelgr}.
Another  distinctive feature of general relativity is the dragging 
of inertial frames due to mass rotation. This fact is
  observed, for instance,  in the impressive
differences of the geodesic motion in Schwarzschild
 and Kerr geometries \cite{chandra}.

Along this article we shall study the geodesic equations for particles evolving
in the space time associated to a center of attraction with a
 quadrupolar deformation. The solution of the Einstein equations 
representing  this  center of attraction   -- in the static case --
 can be found in \cite{z} wherein the rather misleading 
terminology ``distorted black hole''
 was used to refer to such an object.
Examples of static centers of attractions with multipolar deformations 
are: a) A true static black hole (or a dense object)
 surrounded by  a distribution of matter like a ring
or a small disk  formed by counter-rotating matter, i.e., built
by  approximately the same number of  particles
 moving  clockwise as counterclockwise.  Even though, this
 interpretation can be 
seen as a device to have a static stable configuration
there is observational evidence of disks 
made of streams of  rotating and counter-rotating matter \cite{counter}.
b) An axially symmetric static dense object with either polar deformations
or polar jets. In the case a) we  have oblate deformations. Also the polar
deformations of the 
 Sun and the inner planets in the solar  system are oblate.  We have 
prolate deformations in  stars with jets and  in 
some galaxy clusters \cite{cigarshape}.
In the precedent cases,   by adding rotation to the central black hole and 
removing the counter-rotating hypothesis we  can 
have  stationary centers of attraction with multipolar deformations.

Geodesic motions   for axially symmetric spacetimes representing
 core-halos system were studied in \cite{wlapjprl}     for  bounded motion
 and in \cite{moura}
for unbounded motions. The case of a slowly rotating center
 of attraction  
with a dipolar halo was considered in \cite{lwrot}  . The geodesic chaos for a 
disk with a central center of attraction was considered in \cite{saa}.
  A core-halo system with NUT (Newman-Unti-Tamburino) charge was also
 considered \cite{lwnut}. Newtonian \cite{newton}  and pseudo-Newtonian \cite{pseudo} 
counter parts of some of this systems has been also studied.  
In a recent paper -- within the Newtonian realm -- we studied chaotic
 motions of
test-particles orbiting  around a deformed body 
modeled by a monopolar and an internal quadrupolar
term \cite{gueron}.

In this  article we  dwell in  study of geodesic chaos, 
but now related to   internal quadrupole deformations of the attraction center.
Note that halos are external multipolar contributions, their strength
 increases with the distance contrary to the internal ones that decreases
 with the distance. The quadrupolar contribution  usually
represents the major deviation to the spherical symmetry.  Thus,  as  a good
first approximation,  it  can model  most of the deformed sources. 

We shall analyze only bounded motions 
 for  specific choices of energy and
angular momentum and  certain values of quadrupolar strength that we
 believe will cover all the different typical situations. Due to the
symmetry of the problem, one  can reduce the geodesic motion to a
 dynamical system with  two degrees of freedom. For such
cases, the Poincar\'e section method is the most appropriated tool
to   study the geodesics general behavior.

The paper consists of  two main parts. In the first one, Sec. II,  the exact 
solution to the Einstein equations that represents  a static axially symmetric
   source deformed by an internal quadrupole is considered. By using 
 the Poincar\'e section method  we numerically 
study the  geodesic motion of test particles.
For the prolate quadrupolar case we  found chaotic motions contrary to the
oblate case where only regular motion was found.

In the second part, Sec. III, the
  rotation of the attraction center  is considered. We first study 
 the metric that represents a rotating black hole deformed by a quadrupolar
term. This metric is obtained   as a two soliton solution
in the context of Belinsky--Zakharov inverse scattering method
 \cite{belinsky} that  generates  new solutions from  
a known one  (seed solution).  As in the precedent section,
 geodesics  were numerically studied using surfaces of section.
The consideration of different cases leads us to conclude that
the black hole rotation  considerably alters the stability of the system.
 We obtain  that the stability
 depends strongly on the relative direction of the spin of the 
center of attraction and the test particle  angular momentum.
We also found that the rotation does not alter the regularity of geodesic
motions in the oblate case, i.e., the orbits in this  case
 remain regular. We conclude, in Sec. IV, with further 
considerations on the stability of orbits. But, now we employ the 
method of Lyapunov characteristic numbers
following  Benettin et al. \cite{bgs}. 
We also discuss the typical time  to observe instability of  orbits and 
make some final remarks.

\section{ Schwarzschild solution with  quadrupole deformations}

The metric of the spacetime related  to  the  gravitational field of a static
 axially symmetric source is the one associated with the  Weyl line element,
\begin{equation}
ds^{2}=e^{2\psi }dt^{2}-e^{2(\gamma -\psi )}(dz^{2}+dr^{2})-r^{2}e^{-2\psi
}d\varphi ^{2},  \label{weyl}
\end{equation}
where $\phi$ and $\gamma$ are functions of $r$ and $z$ only. The range of the
coordinates $r,z,\varphi$ are the usual ones for cylindrical coordinates.
It is more convenient to use prolate spheroidal coordinates, $u$ and  $v$, that
are related to  Weyl coordinates by 
\begin{eqnarray}
r^{2} &=&m^{2}(u^{2}-1)(1-v^{2}),  \label{transf} \\
z &=&muv,  \nonumber
\end{eqnarray}
where $m$ is a constant, that will be associated with the mass of the center of attraction. The coordinate $v$ takes values in
the interval $[-1,1]$ (it is essentially a cosine) and  $u$ runs
from $1$ to infinity (it is essentially a radial coordinate).
 We shall use
units such that $c=G=1$. With no  lose of 
generality  we shall   also choose  $m=1$.
In this new system of coordinates, the metric (\ref{weyl}) takes the form , 
\begin{eqnarray}
ds^{2} &=&e^{2\psi (u,v)}dt^{2}-e^{-2\psi (u,v)}(u^{2}-1)(1-v^{2})d\phi ^{2}
\nonumber\\
&&-e^{2(\gamma (u,v)-\psi (u,v))}(u^{2}-v^{2})\left( \frac{du^{2}}{u^{2}-1}+%
\frac{dv^{2}}{1-v^{2}}\right) . \label{weylxy}  
\end{eqnarray}

For this line element the vacuum Einstein equations reduce  to,
\begin{equation}
\lbrack (u^{2}-1)\psi ,_{u}],_{u}-[(1-v^{2})\psi ,_{v}],_{u}=0,
\label{laplace}
\end{equation}
\begin{eqnarray}
\gamma _{,u} &=&\frac{(u\psi _{,u}-2v\psi _{,v})(u^{2}-1)(1-v^{2})\psi
_{,u}-u(1-v^{2})^{2}\psi _{,v}{}^{2}}{(u^{2}-v^{2})},  \nonumber \\
\gamma _{,v} &=&\frac{(2u\psi _{,u}-v\psi _{,v})(u^{2}-1)(1-v^{2})\psi
_{,v}+v(u^{2}-1)^{2}\psi _{,u}{}^{2}}{(u^{2}-v^{2})}.  \label{E2}
\end{eqnarray}
Equation (\ref{laplace}) is the usual Laplace equation in prolate
 coordinates for the metric potential $\psi$ and Eqs. (\ref{E2}) yield the
function $\gamma$ as a quadrature. The ingrability of $\gamma\;$  ($\gamma_{,uv}=
\gamma_{,vu}$) is guaranteed by Eq. (\ref{laplace}).  The potential $\psi$ for
 the Schwarzschild solution in prolate coordinates is \cite{W} ,
\begin{equation}
\psi_S =\frac{1}{2}\log  \frac{1-u}{1+u}.
\end{equation}
In this article we shall consider the solution, 
\begin{equation}
\psi =\frac{1}{2}\log  \frac{1-u}{1+u}
+k_{2}P_{2}(v)Q_{2}(u),  \label{psi}
\end{equation}
where $P_2$ and $Q_2$ are the  second Legendre polynomial and  function, 
\begin{equation}
P_2(x)=(3x^2-1)/2, \;\; Q_2(x)=[P_2(x)\log\frac{x+1}{x-1}-3x]/2, \label{leg}
\end{equation}
and $k_2$ is a constant that is positive (negative) for oblate (prolate)
 deformations. Note that the Newtonian limit of the potential (\ref{psi})
 is  \cite{elhers}, $\phi=-m/R+(2m^3k_{2}/15)P_2(\cos\vartheta)R^{-3}$.

From (\ref{E2}) we find the other metric function,
\begin{eqnarray}
\gamma&=&\{4 [2 ((7 k_2^2 - 20 k_2 + 4) \log(u - 1) + (k_2 + 2)^2 \log(u + 1) - 4\log(u^2 - v^2) (k_2 - 1)^2) - 3 ((27 u^2 v^2 - 30 u^2  - 21
v^2 +\nonumber \\
 & &26) k_2 - 8) \log((u + 1)/(u - 1)) k_2 u v^2 + 3 ((27 u^2 v^4 - 30 u^2
v^2 + 3 u^2 - 12 v^4 + 16 v^2) k_2 - 16 v^2) k_2] -\nonumber \\& & 3 [4 ((3 u^2 - 3 u -
2) k_2 + 8) - 3 (9 u^2 v^2 - u^2 - v^2 + 1) (u - 1) (v^2 - 1)\nonumber \\& & \log((u + 1)/(u - 1)) k_2] (u + 1) \log((u + 1)/(u - 1)) k_2\}/64. \label{gam}
\end{eqnarray}

The exact solution to  Einstein equations given by (\ref{psi})-(\ref{gam})
was first study by Erez and Rosen \cite{ER}, we will comeback
 to this point latter. The general case (Schwarzschild 
with the whole series of multipoles) was considered by Quevedo \cite{quevedo}
and a simple interpretation in terms of bars was presented by Letelier 
\cite{letbar}. This solution has been interpreted as a ``distorted''
 black hole
in \cite{z}. The study of the associated 
Newtonian multipoles as well as the relativistic multipoles for this solution and other multipolar expansion can be found in \cite{boisseau}.
The geodesic equations for the metric (\ref{weylxy}) take the form,
\begin{eqnarray}
&&\frac{d^{2}u}{d\tau ^{2}} =\frac{u^{2}-1}{2e^{2(\gamma -\psi
)}(u^{2}-v^{2})} \left( \partial _{u}e^{2\psi }+\partial
_{u}[(u^{2}-1)(1-v^{2})e^{-2\psi }]\right) -\dot{u}^{2}\left( [\partial
_{u}(\gamma -\psi )]+\frac{(v^{2}-1)u}{(u^{2}-v^{2})(u^{2}-1)} \right)
\nonumber \\
&&\hspace{1cm}-2\dot{u}\dot{v}\left( [\partial _{v}(\gamma -\psi )]-\frac{v}{%
(u^{2}-v^{2})} \right) -\dot{v}^{2}\left( \frac{(u^{2}-1)[\partial
_{u}(\gamma -\psi )]}{(v^{2}-1)} +\frac{(u^{2}-1)u}{(u^{2}-v^{2})(v^{2}-1)}
\right) ,\label{geo1} \\
&&\frac{d^{2}v}{d\tau ^{2}} =\frac{1-v^{2}}{2e^{2(\gamma -\psi
)}(u^{2}-v^{2})} \left( \partial _{v}e^{2\psi }+\partial
_{v}[(u^{2}-1)(1-v^{2})e^{-2\psi }]\right) -\dot{v}^{2}\left( [\partial
_{v}(\gamma -\psi )]-\frac{(u^{2}-1)v}{(u^{2}-v^{2})(v^{2}-1)} \right)
\nonumber \\
&&\hspace{1cm}-2\dot{u}\dot{v}\left( [\partial _{u}(\gamma -\psi )]+\frac{u}{%
(u^{2}-v^{2})} \right) -\dot{u}^{2}\left( \frac{(v^{2}-1)[\partial
_{v}(\gamma -\psi )]}{(u^{2}-1)} -\frac{(v^{2}-1)v}{(u^{2}-v^{2})(u^{2}-1)}
\right)\label{geo2} ,\\
&& E =e^{2\psi (u,v)}\dot{t},  \;\; L =e^{-2\psi (u,v))}(u^{2}-1)(1-v^{2})\dot{\varphi},  \label{cte}
\end{eqnarray}
where $\tau=s/c=s$ and the overdots indicate derivative with respect $\tau$.
$E$ and $L$ are constants of integrations related to the test particle
 energy and the angular momentum, respectively. The metric  (\ref{weylxy}) gives a third constant of motion,

\begin{eqnarray}
1 =e^{2\psi (u,v)}\dot{t}^{2}-e^{-2\psi (u,v)}(u^{2}-1)(1-v^{2})\dot{\varphi} ^{2}-e^{2(\gamma (u,v)-\psi (u,v))}(u^{2}-v^{2})\left( \frac{\dot{u}^{2}}{u^{2}-1}+%
\frac{\dot{v}^{2}}{1-v^{2}}\right)\label{constmet}  .  
\end{eqnarray}

The motion of the test particle is completely determined by the solution of the
two second order differential equations (\ref{geo1}) and (\ref{geo2}). They define a four dimensional phase space, but the motion constants
 (\ref{constmet}) and (\ref{cte})
 tell us that the motion is effectively realized in a
 three dimensional surface. Moreover,  these constants allow
us to define an effective potential like function,
\begin{equation}
\Phi (u,v)=\frac{e^{2(\psi (u,v)-v(u,v))}}{(u^{2}-v^{2})}\left( e^{-2\psi
(u,v)}E^{2}-\frac{e^{2(\psi (u,v)-v(u,v))}}{(u^{2}-1)(1-v^{2})}%
L^{2}-1\right) .  \label{efpot}
\end{equation}
Thus the motion must be restricted to the region defined by the
 inequality $\Phi (u,v)\leq0$.

Since the geodesic motion of the test particles is performed in a
 three dimensional effective phase space an adequate tool to study this motion 
is the Poincar\'{e} section method.  As we  mentioned  before 
the sign of the quadrupole parameter  $k_2$  specifies if the source
 is deformed in a prolate or in an oblate
shape. First we shall  study the prolate case, $k_2 < 0$.

From  relation  (\ref{efpot})  we  can find
  the appropriated constants to have  a 
confined  motion. We find that only three combinations of the constants:
 $E$ (energy),  $L$ (angular momentum) and  $k_2$ (quadrupole strength),
 characterize all the possibilities of confinement. In Fig.\ref{csi1} we present the curve $\Phi(u,v)=0$
for $L=3.32$, $E=0.937$ and $k_2=-5.02$.  We have two bounded regions and 
two unbounded ones.
 With the same values to $L$ and $E$ and  a small change in
the quadrupolar constant, $k_2=-5.0$, we obtain
the curve plotted in Fig.\ref{csi2}. The 
two  bounded regions  merge in a single one. The two escape zones  
 remain unbounded. Finally, in figure 3, we
present the curve $\Phi(u,v)=0$ for $L=3.8$, $E=0.9731$ and $k_2=-1$. 
Now the two zones of unbounded motion merge in a single one and the
region of bounded motions increases.     

We construct Poincar\'e  surface sections for the the three sets of constants
indicated above.  In  Fig.\ref{ps1} we present a Poincar\'e section
for the two bounded regions of   Fig. \ref{csi1}. 
 In the middle  bounded region   we have a typical picture
of chaotic motion. However the orbits confined in the right hand side
 bounded region present  regular motion.  In Fig.\ref{ps2} we show
 the Poincar\'e
section obtained for orbits restricted to the closed surface presented 
in Fig.\ref{csi2}. It is interesting to observe that we have
a region of irregular motion  in the left hand side of the graphic
 very similar to the one
showed in the previous figure  and in the right hand 
 side  a region of regular motion
surrounded by a chaotic one. 

In Fig. \ref{ps3} we show that the motion in the bounded region of
 Fig. \ref{csi3} is regular as in the case of a pure
 Schwarzschild black hole \cite{wlapjprl}. 
These results can be understood by studying the effective potential
 critical points.
We recall that a pure   black hole ($k_2=0$) with adequate values
 of the constants $E$ and $L$ has an
effective potential with  a single saddle point. When we add the prolate
 quadrupolar field  $k_2<0$ we have a second saddle point for the value of the
constants of Figs. \ref{csi1} and \ref{csi2}.  In the third case
 (Fig. \ref{csi3}) the second saddle point disappears  and we end up with the
 same dynamical behavior  of the test particles  as in the pure
  Schwarzschild black hole case.

For the case of oblate quadrupole deformation,
i.e., $k_2>0$,  we found  regions of bounded motion very similar
to the one presented in Fig. \ref{csi3}. But, we did not find a second as in
Figs. \ref{csi1} and \ref{csi2}. This indicates  the absence of a 
second saddle point. We studied 
surface sections for many 
different values for $E$, $L$ and $k_2>0$. We always found
 regular motion.
 
\section{Kerr solution with   Quadrupole deformations}

Since the Kerr solution represents a rotating black hole, the addition of an
internal multipole term can be used to model a rotating star or the core of
a galaxy. The black hole rotation produces the pure relativistic effect of
dragging of inertial frames. Then our main goal,  in this section, is to study
the influence of the black hole rotation on the stability of geodesic motions.
Letelier and Vieira \cite{lwrot} studied the motion of test particles
 moving around a slowly rotating black hole with a dipolar halo. 
Now we shall 
study the
case of a central body with  arbitrary rotation  deformed by an internal
 quadrupole term. 

The metric for a stationary axially symmetric spacetime has the general
form, 
\begin{equation}
ds^{2}=g_{tt}dt^{2}+\!2g_{t\phi }dtd\phi +g_{\phi \phi }d\phi
^{2}-e^\Gamma\,(u^{2}-v^{2})\left[ \frac{du^{2}}{u^{2}-1}
+\frac{dv^{2}}{1-v^{2}}%
\right] ,  \label{papa}
\end{equation}
where  $g_{tt}$, $g_{t\phi }$, $g_{\phi \phi }$ and $\Gamma$ are function of
 the coordinates $u,v$.

Belinsky and Zakharov presented a solution generating algorithm for metrics
with two independent Killing vectors \cite{belinsky}. They obtained the Kerr
solution by applying the method to the Minkowski spacetime (seed solution).
The application of this solution generating method to
 more general seeds was studied in \cite
{letbz}. Using the techniques presented in this last article we can
easily obtain the metric functions $g_{tt}(u,v),$ $g_{t\phi }(u,v),$ $%
g_{\phi \phi }(u,v)$ and $f(u,v)$ that represent a Kerr black hole deformed
by  multipolar terms. We choose as the
  seed a metric representing a pure
quadrupole. Then the the two-soliton solutions  give us the 
nonlinear superposition of a Kerr solution with a quadrupolar field. We find,

\begin{eqnarray}
g_{tt}& =& (e^{H} (e^{2 H} ((2 e^{2 F_1+2  F_2}
  (u^2-v^2)-e^{2 H} (v^2-1)) (p+1)^2 -e^{4 F_1} (u+1) (u-1) q^2) q^2 
\nonumber\\
 & &-e^{4 F_2}  (e^{2 H} (p+1)^2 (u+1) (u-1)+e^{4 F_1} (v^2-1) q^2) (p+1)^2))/ \nonumber\\
 & &(e^{2 H} ((2e^{2 F_1+2 F_2} (u+v)(u-v)-e^{2 H} (v-1)^2) (p+1)^2
  e^{4 F_1} (u-1)^2 q^2) q^2-e^{4 F_2}\nonumber \\
 & & (e^{2 H} (p+1)^2 (u+1)^2+e^{4F_1} (v+1)^2 q^2) (p+1)^2), \label{gk1}  
\\
& &\nonumber\\
 g_{t\phi}&=&(-2 e^{H} (e^{2 H} ((2 e^{2 F_1+
      2 F_2} (u^2-v^2) \nonumber  \\ & & -e^{2 H} (v^2-1)) (p+
      1)^2-e^{4 F_1} (u+1) (u-1) q^2) q^2- \nonumber \\ & &e^{4 
      F_2} (e^{2 H} (p+1)^2 (u+1) (u-1)+e^{4 
      F_1} (v^2-1) q^2) (p+1)^2+\nonumber \\ & &(e^{2 F_1} 
      (e^{4 F_2} (p+1)^2 (u+1) (v+1)+e^{2 H} 
      (u-1) (v-1) q^2) (u-v)- \nonumber \\ & & e^{2 F_2} (e^{2 H}
       (p+1)^2 (u+1) (v-1)+e^{4 F_1} (u-1) (v+1) 
      q^2) (u+v)) (p+1) p) q)/ \nonumber \\  & & (e^{2 H} ((2 e^{2 F_1+2 F_2} (u^2-v^2)-e^{2 H} (v-1)^2) (p+
      1)^2-e^{4 F_1} (u-1)^2 q^2) q^2-\nonumber \\ & & e^{4 F_2
       } (e^{2 H} (p+1)^2 (u+1)^2+e^{4 F_1} (
      v+1)^2 q^2) (p+1)^2), \label{gk2}\\
 & &   \nonumber \\ 
    g_{\phi\phi}& =&\frac{g_{t\phi}^2-p^2(1-v^2)(u^2-1)}{g_{tt},}\label{gk3}\\
 & &   \nonumber \\
 e^\Gamma & = &-(\exp[((((4  (2  \log(u+1 )+81  u^2 v^4-90  u^2 v^2+
       9  u^2-36  v^4+48  v^2-8  \log(u+v)+\nonumber\\
   & &14  \log(u-1 )-8  
       \log(u-v))+9  (9  u^2 v^2-u^2-v^2+1 ) (u^2-1) (v^2-1)
        \log((u+1 )/(u-1 ))^2\nonumber\\ 
& &-12  (27  u^3 v^4-
       30  u^3 v^2+3  u^3-2 -(21  v^2-5 ) (v^2-1) u)
        \log((u+1 )/(u-1 ))) k_2-\nonumber \\ 
& & 16  ((3  u^2-1 ) \log((u+1 )/(u-
       1 ))-6  u) (3  v^2-1 )) k_2)/128 )] (e^{2  H} ((
       2  e^{2  F_1+2  F_2} (u^2-v^2)-\nonumber\\
& &e^{2  H} 
       (v-1 )^2) (p+1 )+e^{4  F_1} (p-1 ) (u-1 )^2) (p-
       1 )+e^{4  F_2} (e^{2  H} (p+1 ) (u+1 )^2-e^
       {4  F_1} \nonumber\\
& &(p-1 ) (v+1 )^2) (p+1)))/(4  e^{2  F_1
       +2  F_2+2  H} (u^2-v^2) p^2),\label{fk} 
\end{eqnarray}
where
\begin{eqnarray}
 F_1&=&(-(2 (2 (\log(u+1)-3 v^2-3 \log(u-1)+2 \log(u-v
      ))+3 (3 v-1) (v+1) u)\nonumber\\
 & &+3 ((v+3+2 (v+1) u) (v-
      1)-(3 v-1) (v+1) u^2) \log((u+1)/(u-1))) k_2)/16, \label{F1}\\
   F_2 & = & (-(2 (3 ((3 v-1) (v+1) u+2 v^2)-4 \log(u+v)+
      4 \log(u-1))-\nonumber \\ 
& & 3 ((3 v-1) u-(v+1)) (u+1) (v+1) 
      \log((u+1)/(u-1))) k_2)/16 \label{F2}\\
          H & = & (((3 u^2-1) \log((u+1)/(u-1))-6 u) (3 v^2-1) k_2)/8 \label{H}
       k_2)/8.
\end{eqnarray}
The quadrupole strength is $k_2$, $q$ is the source angular momentum 
per the square of the mass and $p$ is defined by the relation $p^2+q^2=1$.
The metric presented above is a particular case of the general solution
that represent a Kerr metric embedded in a field of multipoles, see 
for instance \cite{quevedo2} and \cite{letbar}, we will back to 
this point latter. 

When one performs the limit $k_2\rightarrow 0$ in the solution presented
 above one obtains the Kerr metric
in Boyer Lindquist coordinates, $r$ and $\vartheta $ that are related to the
prolate spheroidal coordinates, $u$ and $v$ by 
$u =(r-m)/\sigma$  and $v =\cos \theta$. The constant $p$ and $q$ are
 related to the Boyer Lindquist constants by $p =\sigma/m, \;\;$ $
q =a/m,$ and 
$m^{2} = \sigma ^{2}+a^{2}.$
$m$ is the monopole mass, $\sigma $ is only an auxiliary constant and $a$ is
 interpreted as the black hole angular momentum per unit of mass
measured by a very distant observer.

As in the precedent case, the geodesic equations have two constants of
motion, $L$ and $E$, that obey the relation.
\begin{equation}
g_{tt}\dot{t}+g_{t\phi }\dot{\phi} =E, \;\;
g_{\phi \phi }\dot{\phi}+g_{t\phi }\dot{t} =L.  \label{geokc}
\end{equation}

The evolution equations for $u$ and $v$ are
\begin{eqnarray}
\frac{d^{2}u}{d\tau ^{2}} &=&\frac{u^{2}-1}{2\Gamma (u^{2}-v^{2})} \left[
\partial _{u}g_{tt}\dot{t}^{2}+2\partial _{u}g_{t\phi }\dot{t}\dot{\phi}
+\partial _{u}g_{\phi \phi }\dot{\phi}^{2}\right] -\dot{u}^{2}\left[ \frac{
\partial _{u}\Gamma }{2\Gamma } +\frac{(v^{2}-1)u}{(u^{2}-v^{2})(u^{2}-1)}
\right] \nonumber \\
& &-2\dot{u}\dot{v}\left[ \frac{\partial _{v}\Gamma }{2\Gamma } -\frac{v}{%
(u^{2}-v^{2})} \right] -\dot{v}^{2}\left[ \frac{\partial _{u}\Gamma
(u^{2}-1)}{2\Gamma (v^{2}-1)} +\frac{(u^{2}-1)u}{(u^{2}-v^{2})(v^{2}-1)}
\right] , \label{geok1}\\
\frac{d^{2}v}{d\tau ^{2}} &=&\frac{1-v^{2}}{2\Gamma (u^{2}-v^{2})}\left[
\partial _{v}g_{tt}\dot{t}^{2}+2\partial _{v}g_{t\phi }\dot{t}\dot{\phi}
+\partial _{v}g_{\phi \phi }\dot{\phi}^{2}\right] -\dot{v}^{2}\left[ \frac{
\partial _{v}\Gamma }{2\Gamma }-\frac{(u^{2}-1)v}{(u^{2}-v^{2})(v^{2}-1)}%
\right] \nonumber \\ 
&&-2\dot{u}\dot{v}\left[ \frac{\partial _{u}\Gamma }{2\Gamma }+\frac{u}{%
(u^{2}-v^{2})}\right] -\dot{u}^{2}\left[ \frac{\partial _{v}\Gamma (v^{2}-1)%
}{2\Gamma (u^{2}-1)}-\frac{(v^{2}-1)v}{(u^{2}-v^{2})(u^{2}-1)}\right]
 \label{geok2} .
\end{eqnarray}

As in the precedent case we have two second order evolution  equations 
(\ref{geok1}) and (\ref{geok2}) for the variables $u$ and $v$, the metric and 
the constants (\ref{geokc}) give a new constant relating these two variables,
\begin{equation}
1=g^{tt}E^{2}+\!2g^{t\phi }EL +g^{\phi \phi }E
^{2}-e^\Gamma\,(u^{2}-v^{2})\left[ \frac{\dot{u}^{2}}{u^{2}-1}
+\frac{\dot{v}^{2}}{1-v^{2}}%
\right] .  \label{metrck}
\end{equation}
In other words, despite  algebraic complications, we have exactly the
 same dynamical situation as before. The particles move in an effective
  three
 dimensional space. Thus  we shall analyze the motion of test particles 
moving in the gravitational field  of 
a rotating  prolate deformed body using Poincar\'e sections as in
 the non-rotating
case.

 Since the main new ingredient in the new system is the rotation of
 the source we shall keep the angular momentum $L$, the energy $E$ and the
 quadrupole strength $k_2$ fixed and we shall consider 
 test particles moving with  angular momentum  parallel to the spin
 source (co-rotation) and with angular momentum anti-parallel to
 the spin source (counter-rotation).

In Fig.\ref{ck1} we present   the region of bounded motions 
for counter-rotating orbits, $qL<0$. We take $E=0.93715,\;\; |L|= 3.322,$
  and $k_2=-5.08$, and for the rotation
 parameter, $q=0.002$. 
 We  notice a situation
similar to the one presented in Fig.\ref{csi2}. 
The bounded regions for the co-rotation case, $qL>0 $ is 
shown in Fig.\ref{ck2}. We see two relatively small and distant closed 
surfaces. The Poincar\'e section for the
counter-rotating case is presented in Fig.\ref{pk1}.
 Chaotic motion can be perceived in the
left hand side of the graphic and in the external part of the right hand
 side  as  in Fig.\ref{ps2}.
In Fig.\ref{pk2}  we present the surface  section for the same 
parameters as above, but  $qL>0$ (co-rotation).  We do not
find  chaotic motion in this case. 

We were not able to obtain bounded motion for both large  quadrupole strength
and  large rotation parameter. We studied bounded systems with large rotation speed
(of the order of 0.1) but with quadrupole strength always less than unit. In
these cases the study of Poincar\'e sections leads
 to regular geodesic motion for co-rotation, as well as, counter-rotation.
 We also found  that the confinement region may suffer an appreciable 
change in size and shape. 

\section{Discussion}

As we said before, the exact solution to  Einstein equations presented in
 the two precedent sections are not new  and different versions of them have 
already appear  in the literature. We have presented them  in this work for two
reasons: a) For easy reference, and b) Mainly, because for
 numerical analysis we need a faultless solution. The one presented here
were derived using algebraic computation and checked using the full 
vacuum Einstein equations in each case.

Besides the Poincar\'e section, we have another technique to
 quantify  geodesic
instability: the Lyapunov characteristic number, $\cal N$, defined as,  
 \begin{equation}
{\cal N}=\lim_{\scriptsize
\begin{array}{l} \delta_0 \rightarrow 0 \\
 \tau\rightarrow \infty  \end{array}}
 \left[\log(\delta /\delta _{0}) \over \tau \right] , \label{lcn}
\end{equation}
 where $\delta _{0}$ and  $\delta $\ are the deviation of
two nearby orbits at times $0$ \ and $t$\, respectively.  Using the
 technique suggested by Benettin et
al. \cite{bgs} -- who studied numerical problems in the computation 
of Lyapunov exponents and Kolmogorov entropy -- one  can get
the largest ${\cal N}$.

Using the constants that define the bound region presented in Fig. \ref{csi1} 
we obtain 
${\cal N}=(9.0\pm 1.0)\times 10^{-17}$, where the maximum ${\cal N}$ was obtained for $u=2.6$, $v=0$ and $p_v=0$. In the right region plotted in Fig. \ref{ps1} we get always
${\cal N}<5\times 10^{-18}$. This  shows, as expected, that the Lyapunov
coefficient in the regions with chaotic motion does not vanish while
in the region with irregular motion it could vanish. Nevertheless, the value of
the coefficient for chaotic motions is very small, it means that the dispersion
of the orbits is slow if compared with local fluctuations in 
the mean potential. 

We found chaotic geodesic motions for the system black hole plus
 internal quadrupole  for a very small range of parameters,
 specifically when there is a second bounded region. When one 
considers a rotating source  one just re-obtained the same behavior
studied in  \cite{lwrot} for rotating centers of attraction with halos. 
The orbits of counter-rotating particles are more unstable that the orbits
for corotating particles.  In reference \cite{lwrot} the case of slow
rotation was considered.

 We also  studied some bounded chaotic motion for large rotation 
speed ($q>0.1$) in
 a Kerr halo system from an exact solution that represents an external 
dipole plus a Kerr black hole. Using the techniques presented in this 
paper we  conclude that the irregularity introduced by external
 multipole terms are much larger than the ones introduced by internal
 multipole terms. We found also  that chaotic motions for large
 rotation speed  are more frequently, i.e., we have irregular motion
 for a  larger range of the constants $E$ and $L$.

It is not easy to predict the role  that chaos could
 have  in measurable
characteristics
 of galaxies. Let us  choose a parameter $\cal T $ as the time to
characterize the dynamics of the system  e.g., to draw the sections
of an invariant
 torus in a regular motion  or a chaotic region in the Poincar\'{e}
section for irregular motion.
 We shall consider that this is the minimum time to
have observable effects.

Rearranging the units to observable parameters, we obtain $\cal T $ in years
from the expression 
\[
{\cal T} =N\times M\times 1.6\times 10^{-13}yr,
\]
where $N$ is the number of steps in the simulations and $M$ is the mass of the
central black hole (in solar masses). 
The step $N$ varies for different systems. It is
 about  $10^{4}$ for a black hole plus halo system
 and about  $10^{16}$ for a black hole plus internal quadrupole system. 
 For a  typical galactic bulge we have  $M=10^{12}M_{\odot }. $ 
Then  for black hole plus halo system we have
${\cal T} =1000yr$ which is a very small value when compared with 
 galactic ages and for the black hole plus internal quadrupole system,
  ${\cal T} = 10^{15} yr$ which is a very large value compared with the 
Universe age ($10^{10} yr$).
Consequently, chaotic relativistic effect may show up in the formation of
structures in a  black hole plus external halo system, work along this
 line will soon be reported.
 The relativistic effect due to the rotation of
 the source may be important.  The  internal
 deformations do not have a  significant contribution in this case. 
Another possible observational 
manifestation of chaos was studied  in \cite{chaos3}

\vspace{0.5cm}
{\em Acknowledgements.}  The authors thank  FAPESP for financial support.
PSL also thanks CNPq.

\begin{twocolumn}   
\begin{figure}[tbp]
\epsfig{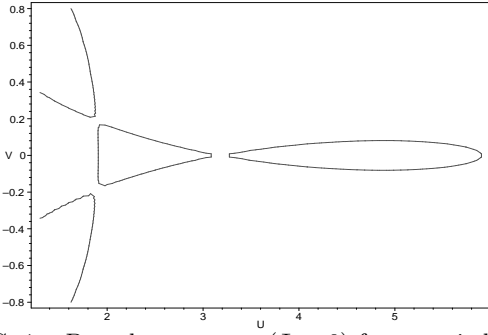}
\caption{ Boundary  contour ($\Phi=0$)
 for a static black hole + quadrupole system  $L=3.32$, $E=0.937$ and $k_2=-5.02$. There are two escape zones in the left hand side of the picture that corresponds to small values of $u$ and a couple of
 closed zones of bounded motion.}\label{csi1}
\end{figure}
\begin{figure}[tbp]
\epsfig{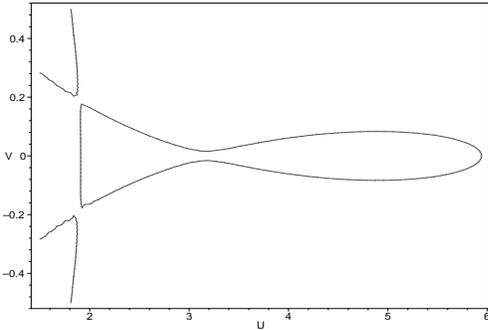}  
\caption{Boundary of contour for the same values for $L$ and $E$
 but $k_2=-5.0$. The two escape  zones remain.  However,  the two closed
 zones  merge in only one. }\label{csi2}
\end{figure}
\begin{figure}[tbp]
\epsfig{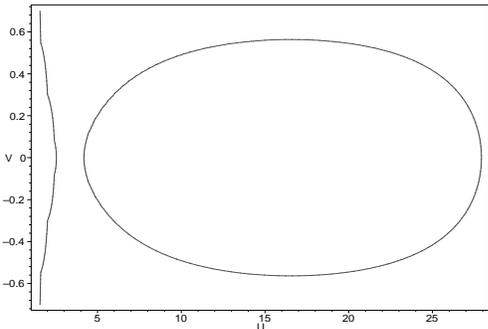}  
\caption{Boundary contour for $L=3.8$ and $E=0.973$ but $k_2=-1.0$. 
No vestige of the second bounded region is left.}\label{csi3}
\end{figure}
\begin{figure}[tbp]
\epsfig{width=3in,height=2in, file=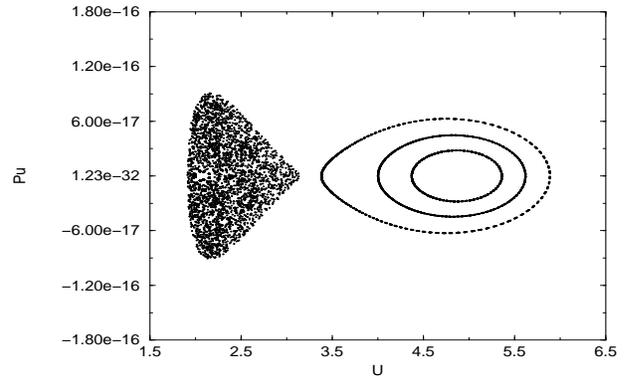}  
\caption{ Poincar\'e Section for the values defined in Fig.\ref{csi1}. We see
  chaotic behavior in  orbits confined in the first zone of bounded motions.
 But,  the motion in the second zone is regular}\label{ps1}
\end{figure}
\begin{figure}[tbp]
\epsfig{width=3in,height=2in, file=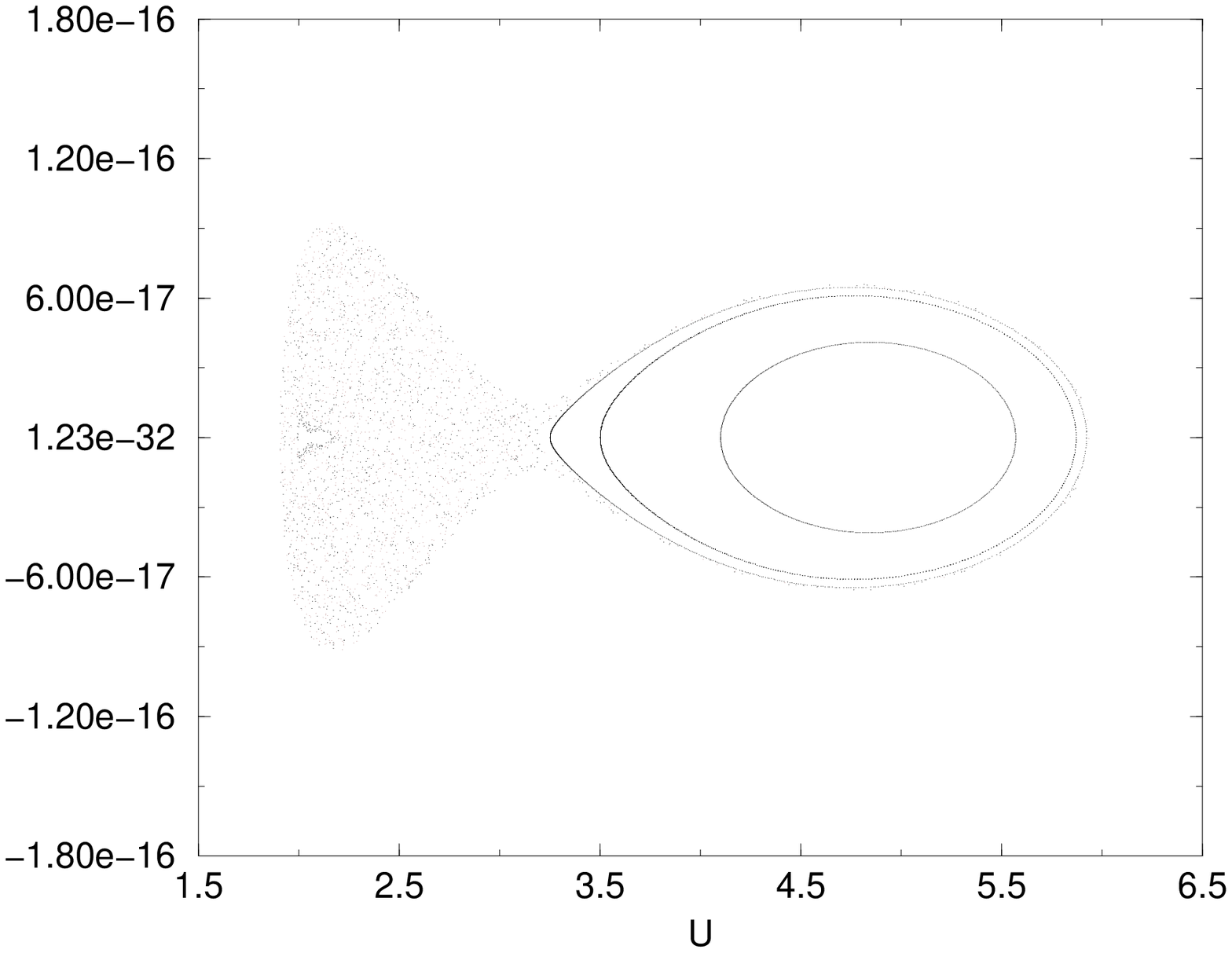}  
\caption{Poincar\'e Section for the values defined in Fig.\ref{csi2}. 
We see  chaotic motion   in the left hand  side of the figure
 and in a small external region of  the right hand side.}\label{ps2}
\end{figure}
\begin{figure}[tbp]
\epsfig{width=3in,height=2in, file=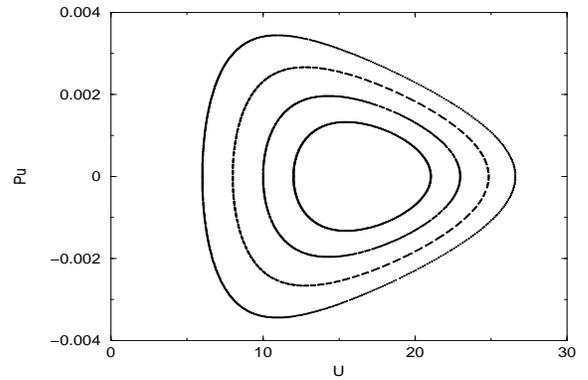}  
\caption{ Poincar\'e Section for the values defined in Fig.\ref{csi3}. We
 have  regular motion. }\label{ps3}
\end{figure}
\begin{figure}[tbp]
\epsfig{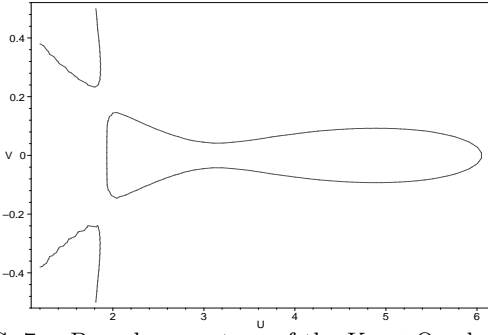}  
\caption{ Boundary contour of the Kerr+Quadrupole system for $L=-3.322$, $E=0.93715$, $k_2=-5.08$ and $q=0.002$. }\label{ck1}
\end{figure}
\begin{figure}[tbp]
\epsfig{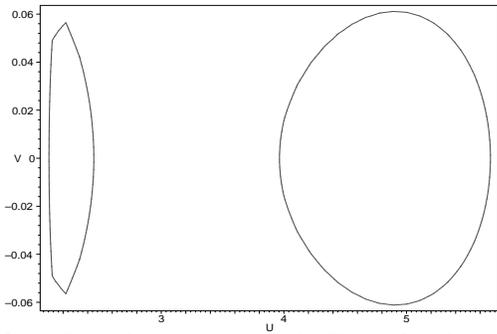}  
\caption{Boundary contour of the Kerr+Quadrupole system for $L=3.322$, $E=0.93715$, $k_2 =-5.08$ and $q=0.002$. The confinement region is
 separated in two. They are much smaller than in the precedent figure}\label{ck2}
\end{figure}
\newpage
\begin{figure}[]
\epsfig{width=3in,height=2in, file=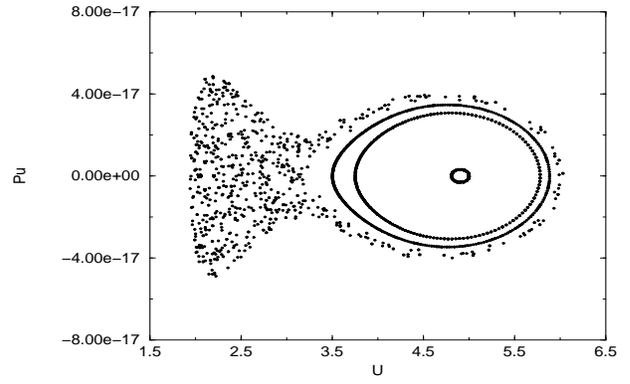}  
\caption{ Poincar\'e Section for the same value of the  parameters
of    Fig. \ref{ck1}. We have 
chaotic
 motion   mainly in the left hand side of the picture}\label{pk1}
\end{figure}
\begin{figure}[]
\epsfig{width=3in,height=2in, file=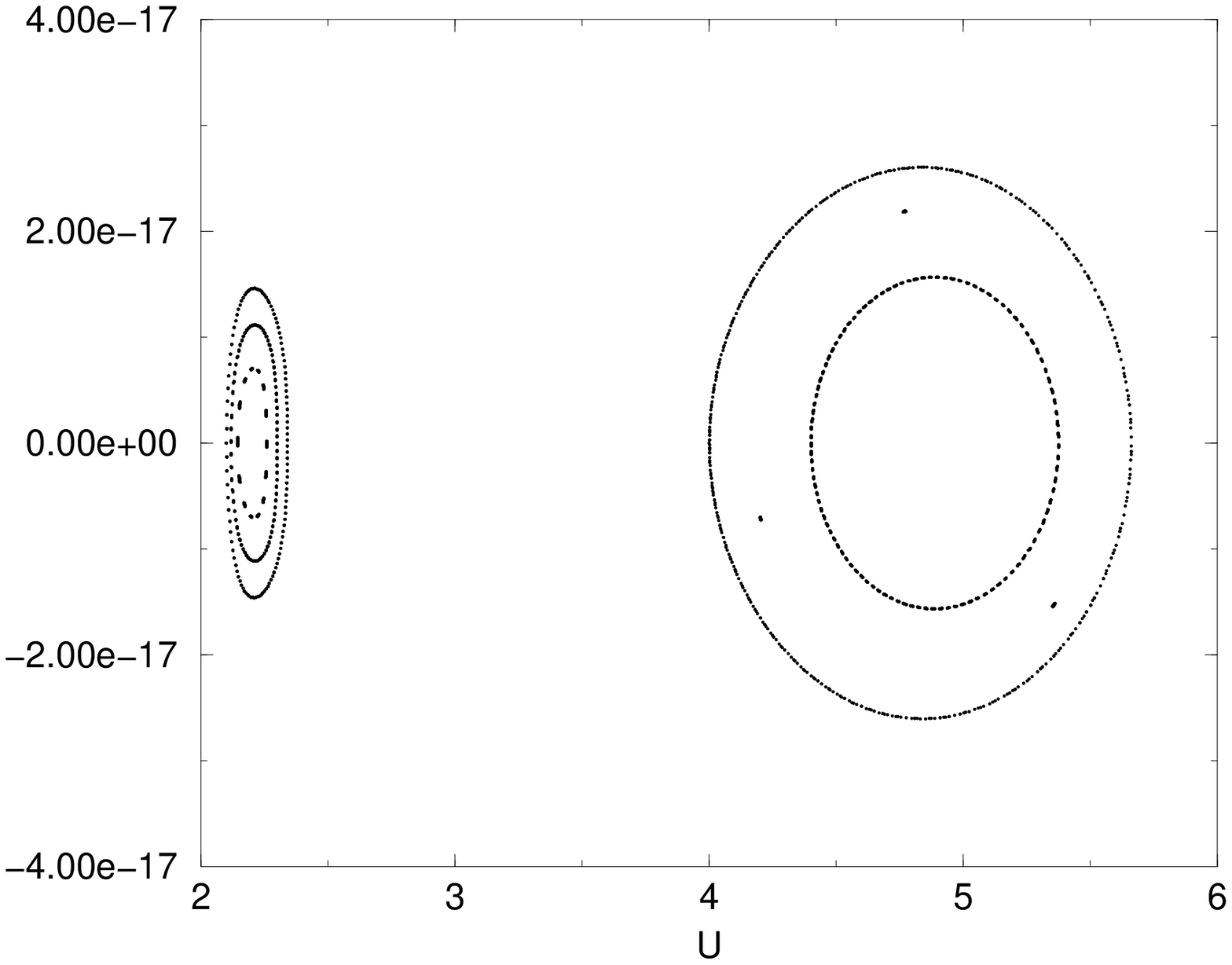}  
\caption{ Poincar\'e Section for the same value of the parameters
of   Fig. \ref{ck2}. We
 do not see  irregular  motion in both   regions}\label{pk2}
\end{figure}

\end{twocolumn}

\begin{references}
\bibitem{poincare}  H. Poincar\'{e} {\it Les M\'{e}thodes Nouvelles de la
Mechanique Celeste,} (Gauthier-Villars, Paris, 1892)
\bibitem{kam} A.N. Kolmogorov,  Dokl. Akad. Nauk. SSSR, 98, 527 (1954).
V.I. Arnol'd., Russ. Math. Surv. 18, 9 (1963); ibid 18, 85 (1963).
J. Moser,Math. Ann., 169, 136 (1967)
\bibitem{berry}
M.V. Berry,  Am. Inst. Phys. Conf. Proc. 46, 16 (1978)
\bb{merr} D. Merrit, Science 271, 337 (1996)
\bb{W} See for instance, H. Robertson and T. Noonan, 
``Relativity and Cosmology"(Saunders, London 1968) pp  272-278. 
\bb{cosg} See for instance,  C.M. Cosgrove, J. Math. Phys. 23, 615 (1982)
 and references therein.
\bibitem{cont} G. Contopoulos, Proc. R. Soc. Lond. A 431, 183 
(1990); A 435, 551 (1991).
\bibitem{ssmae} Y. Sota, S. Suzuki and K. Maeda, Class. Quantum 
Grav. 13, 1241 (1996). See also, W.M. Vieira and P.S. Letelier,
Class. Quantum Grav. 13, 3115 (1996).
\bibitem{wldi} W.M. Vieira and P.S. Letelier, Phys. Lett. A, 288, 22 (1997).
\bibitem{bc} L. Bombelli and E. Calzetta, Class. Quantum Grav. 
9, 2573 (1992).
\bibitem{lwmelgr} P.S. Letelier and  W.M. Vieira, Class. and Quantum Grav.,
14, 1249 (1997).
 \bibitem{chandra} S. Chandrasekhar, ``The mathematical theory of
 black holes", (Clarendon Press, Oxford, 1983).
\bb{z} See for instance, Ya. B. Zeldovich and I.D. Novikov, ``Relativistic Astrophysics"  (University of Chicago Press, Chicago 1971) pp 130-134; M. Carmeli, ``Classical Fields: General Relativity and Gauge Theory" (John Wiley, New York 1982) pp 177-182.
\bb{counter} F. Bertola {\it et al.},
Ap. J.  458, L67 (1996).
\bibitem{cigarshape} A. R. Cooray, Mon Not. R. Astron. Soc. {\bf 313}, 783
  (2000).
\bb{wlapjprl}  W. M. Vieira and P. S. Letelier, Ap. J  513, 383 (1999);
Phys. Rev. Lett. 76, 1409 (1996).
\bb{moura} A.P.S. Moura and P.S. Letelier,
Phys. Rev. E, 61, 6506 (2000). 
\bb{lwrot} P.S. Letelier and W.M. Vieira,  Phys.  Rev.
 D  56,8098 (1997).
\bibitem{saa}  A. Saa and R. Venegeroles, Phys Lett A {259}, 201 (1999); 
see also, A. Saa, Phys Lett A {\bf 269}, 204 (1999).
\bb{lwnut} P.S. Letelier and W.M. Vieira, 
Phys.  Lett.  A,   244, 324 (1998).
\bb{newton}  A.P.S. Moura and P.S. Letelier,
  Physics Letters A, 266, 309 (2000);
  P.S. Letelier and W.M. Vieira, Phys.  Lett. A, 242, 7 (1998).
 P.S. Letelier and A.E. Motter, 
Phys. Rev. E 60, 3920  (1999).
\bb{pseudo} E. Gueron and P.S. Letelier Astron.
 Astrophys. 368, 716 (2001);
 H. Varvoglis and D. Papadopoulos, 
Astron. Astrophys.  261, 664 (1992);
  V. Karas and D. Vokrouhlicky, Gen. Rel.  Gravit. 
 24, 729 (1992).
\bibitem{gueron}  E. Gueron and P.S. Letelier, Phys. Rev. E  63 , 035201(R) 
(2001).
\bibitem{belinsky}  V. A. Belinsky and V. Zakharov, Sov Phys. JETP {\bf 50},
1 (1979)
\bibitem{bgs} G. Benettin, L. Galgani and JM. Streclyn, Phys. Rev. A {\bf 14}
  2338 (1976) 
\bb{elhers} J. Ehlers, in ``Grundlagenprobleme der Modernen Physik'', A. Erd\'elyi, J.Pfarr, and E.-W. Stachov, Eds. (BI-Verlag, Mannheim, 1981) pp 65-84. 
\bb{ER} G. Erez and N. Rosen, Bull. Res. Counc. Isr.  8F, 47 (1959).
\bibitem{quevedo}  H. Quevedo, Phys Rev. D {\bf 39 }2904 (1989)
\bb{letbar}P.S. Letelier, Class. Q.  Grav.,  16, 1207 (1999).

\bibitem{boisseau} B. Boisseau and P.S. Letelier ``
 Relativistic Multipoles and the  Advance of the
Perihelia'',
 preprint Universit\'e de Tours (2001).
\bibitem{letbz}  P. S. Letelier, J. Math Phys.  26, 467 (1984)
\bibitem{quevedo2}  H. Quevedo and B. Mashhoon, Phys Rev. D  43, 3902 (1991)
\bibitem{chaos3}   J. Levin, Phys. Rev. D {\bf 60}, 4015 (1999)



\end{references}
\end{document}